\documentclass[10pt]{NSP1}
\usepackage{url,floatflt}
\usepackage{helvet,times}
\usepackage{psfig,graphics}
\usepackage{mathptmx,amsmath,amssymb,bm}
\usepackage{float}
\usepackage[bf,hypcap]{caption}
\usepackage{algorithmic}
\usepackage{algorithm}
\usepackage{array}

\emergencystretch=\maxdimen
\def\no{\noindent}

\def\firstpage{113}
\def\thevol{9}
\def\thenumber{2}
\def\theyear{2020}
\begin{document}

\titlefigurecaption{{\large \bf \rm Information Sciences Letters }\\ {\it\small An International Journal}}

\title{GASDUINO-Wireless Air Quality Monitoring System Using Internet of Things}

\author{M. E. Karar\hyperlink{author1}{$^{1,2,\star}$}, A. M. Al-Masaad\hyperlink{author3}{$^{1}$} and Omar Reyad\hyperlink{author4}{$^{1,3}$}}
\institute{$^1$College of Computing and Information Technology, Shaqra University, Shaqra, KSA \\
           $^2$Faculty of Electronic Engineering, Menoufia University, Menoufia, Egypt\\
           $^3$Faculty of Science, Sohag University, Sohag, Egypt}

\titlerunning{GASDUINO-Wireless Air Quality Monitoring System Using Internet of Things}
\authorrunning{M. E. Karar et al.}

\mail{mkarar@su.edu.sa}

\received{23 Feb. 2020}
\revised{19 April 2020}
\accepted{23 April 2020}
\published{1 May 2020}

\abstracttext{The Health Effects Institute (HEI) reported recently that the deaths from the negative health effects of the air pollution in the Middle East Region is about 500,000 people. Therefore, this paper presents a new design and development of portable system; called GASDUINO that allows the user to measure the quality of air using the Internet of Things (IoT). The main components of developed GASDUINO system are the Arduino microcontroller board, Gas sensor (MQ-135), Android user interface (UI) connected with all things via Remote XY Arduino cloud. The developed system can alarm the users about the dangerous levels of the air quality index (AQI) or the particle per million (PPM) levels in the range of 0 to above 200 PPM. The developed GASDUINO system is considered as an essential environmental module in the development and sustainability of future smart cities.}


\maketitle

\section{Introduction}  \label{intro}
\begin{figure*}[h]
	\centering
	\includegraphics[width= 0.75\textwidth]{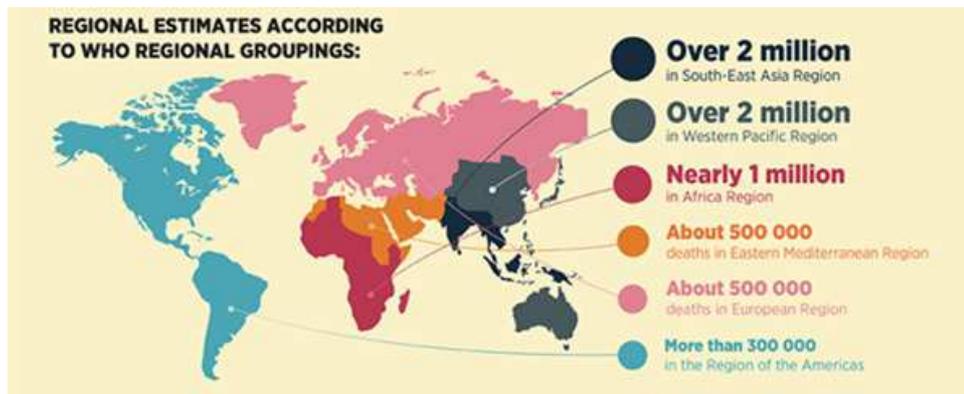}
	\caption{Statistics of the World Health Organization (WHO) for the global deaths due to the air pollution around the world.}
	\label{Fig.1}       
\end{figure*}

Air pollution represents the contaminations in the air including harmful solid particles \cite{c1}. The suspended particles in the air can be car emissions, dust, and poisoned chemical from factories \cite{c2}. Negative effects of air pollution on the human health may cause chronic diseases, for example, asthma, cardiovascular diseases, brain and lung disorders \cite{c3}. The death rate from air pollution in Saudi Arabia is about 30 people per 100,000 people, and globally about 500,000 deaths in Eastern Meditterian Region based on the reports of the Health Effects Institute (HEI) and the World Health Organization (WHO), as shown in Figure \ref{Fig.1} \cite{c4}. The impact of high levels of air pollution can harm the humans as a silent killer causing a brain damage and death eventually \cite{c5}. 

Internet of Things (IoT) presents a good solution to collect remotely the measured levels of harmful gases emission and polluted air index in many different areas of big cities and industrial regions \cite{c6}. The IoT technology can be easily integrated with artificial intelligence, embedded systems, and mobile applications to monitor and analyze big data of the air quality index measurements in the developed framework of smart cities and also in smart homes \cite{c7}. Furthermore, the IoT application to air quality monitoring demonstrated the possible benefits of long-term, cost-effective solution, real-time capabilities, and reporting the air quality index (AQI) information to contribute for the concept definition of Industry 4.0 \cite{c8,c9}. Moreover, security and privacy techniques \cite{c10} can be implied in the proposed IoT systems to provide restrictions on the AQI information access for all stakeholders and industrial organizations \cite{c11}.

\begin{figure*}[h]
	\centering
	\includegraphics[width= 0.75\textwidth]{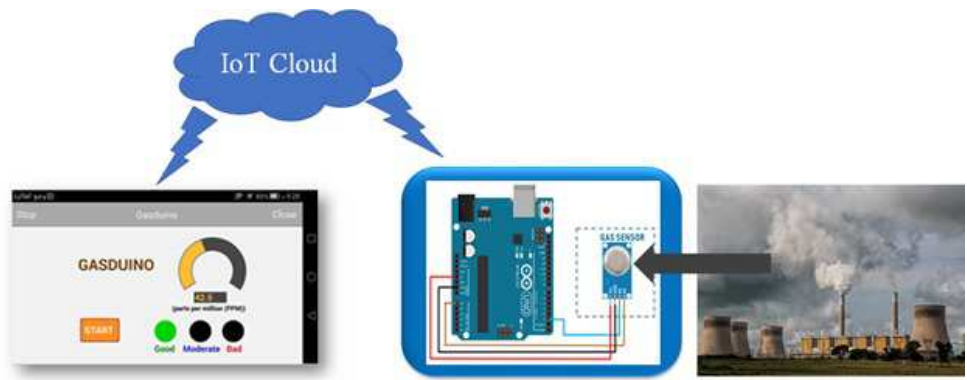}
	\caption{Schematic diagram of the proposed GASDUINO system including the Arduino microcontroller, MQ-135 Gas sensor, and the IoT cloud via Android user interface (UI).}
	\label{Fig.2}       
\end{figure*}

In the previous studies, many IoT applications have been proposed for monitoring the values of AQI either indoor or outdoor as follow. Tudose et al. \cite{c12} developed a prototype of mobile system to measure the air pollution in crowded urban areas, based on wireless sensor nodes (WSN) with database and web server. The mobile unit can be embedded in a car to use its power supply. It is based on an AVR microcontroller to read the gas sensors and connected to LCD for monitoring the instantaneous values of pollutants. In addition, the General Packet Radio Service Modem (GPRS)/Global Positioning System Module (GPS) modem was used to collect these measurements and send them to the web server. However, this proposed air monitoring system showed complexity to be validated with relatively high cost. Using the benefits of google maps and on-line GPRS/GPS module, Al-Ali et al. \cite{c13} proposed air monitoring system including a Mobile Data-Acquisition Unit (Mobile-DAQ), an array of gas sensors, and a pollution web server. Also, Patil et al. \cite{c14} proposed a system based on the WSN to detect various gases to estimate the AQI with ZigBee wireless communication module. Recently, Gokul et al. \cite{c15} exploited the utilities of IoT technology for developing air quality monitoring and forecasting system in Coimbatore as example of smart city in India. They developed an open access website to display and save the measured values of classified AQI. Nevertheless, all previous studies have a lack to introduce a practical mobile application to accomplish the monitoring and classification of instantaneous measured values of AQI.

Therefore, it is necessary to develop a new advanced tool to measure the quality of the air anywhere. In the framework of digitalization and vision 2030, this project uses the technology of IoT to remotely monitor and classify the levels of air quality in the cities of Saudi Arabia, especially in industrial regions. Hence, this project aims at developing an advanced and portable system that can measure the air pollution using IoT embedded system.

The reminder of this article is organized as follows. Section \ref{Sec1} describes the tools and proposed system design that are required to implement our developed air quality monitoring; namely GASDUINO. Section \ref{Sec2} presents practical results and validation of the developed GASDUINO system. Finally, this study is concluded with remarks of our future work in Section \ref{Sec3}. 

{\begin{table*}[h]
		\caption{Categories of the Air Quality Index (AQI) levels in this study.}
		\label{Tab.1}
		\begin{center}
			\begin{tabular}{| m{3cm} | m{4cm} | m{4cm} |} \hline 
				\textbf{Air Quality Status} & \textbf{Air Quality Index- Parts per Million (PPM)} & \textbf{Description}  \\  \hline
				Good         & 0 to 50     & Air pollution is little with no risk. The cleanness of the air is satisfied.   \\ \hline
				Moderate     & 51 to 150   & Accepted quality of the air conditions, but some pollutants may cause diseases for sensitive people.    \\ \hline
				Unhealthy    & 151 to 200  & Very harmful and may cause death.    \\ \hline
			\end{tabular} 
		\end{center}
\end{table*}} 

\begin{figure*}[h]
	\centering
	\includegraphics[width= 0.80\textwidth]{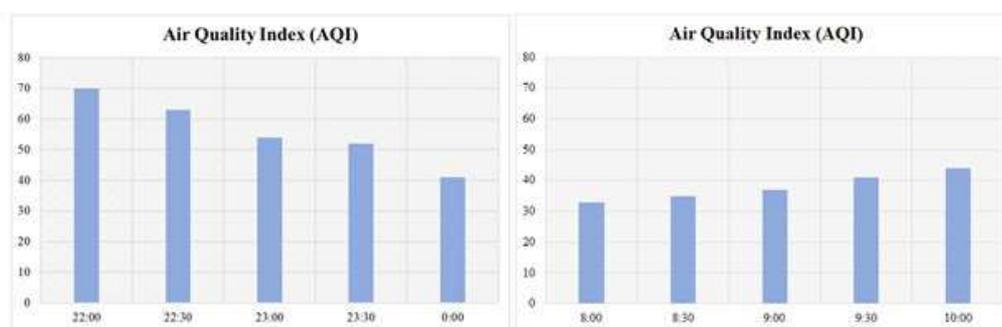}
	\caption{Tested measurements of the air quality index using the gas sensor MQ-135 for two hours during the night (\textbf{\textit{left}}) and day (\textbf{\textit{right}}) in Shaqra city.}
	\label{Fig.3}       
\end{figure*}

\section{Materials and Methods} \label{Sec1}
\subsection{Hardware and Software Tools} \label{Sec1.1}
In this paper, the Arduino UNO board was used as a compatible microcontroller board to connect all other hardware modules of the GADUINO system. Wi-Fi Module (ESP8266) is used to transmit the data via the Internet in the IoT cloud environment. Gas Sensor MQ-135 is an air quality sensor for detecting a wide range of gases, including NH3, NOx, alcohol, benzene, smoke, and CO2. For software Tools, Remote XY Application has been used to develop the mobile application user interface (UI) for handling the Arduino hardware components, based on the Android Mobile operating System. Arduino IDE was used to program the microcontroller with gas sensor MQ-135.

\subsection{Proposed Air Pollution Monitoring System} \label{Sec1.2}
The term "Air quality" represents the cleanness definition of the air with respect to the pollution level. Hence, high pollution levels mean low air quality and lead to cancerous diseases and death as described above, especially in chemical factories and big cities with high population density \cite{c5}. The AQI is the most popular index of the air quality to convert technical air pollution information into public understanding \cite{c16}. The AQI was established in the year $1999$ by the United States Environmental Protection Agency \cite{c17}. It is classified into different categories to show the polluted level of air, ranging from good to unhealthy cases, as illustrated in Table \ref{Tab.1}.

Figure \ref{Fig.2} depicts the proposed GASDUINO system to monitor and classify air pollution levels using Arduino board with a gas sensor (MQ-135) and Android mobile application to display the instantaneous level of the air quality. The IF-Then rule has been applied to identify the air quality status (see Table \ref{Tab.1}). Green, blue, and red indicators represent good, moderate, and unhealthy (or bad) conditions of the air, respectively.

\begin{figure*}[h]
	\centering
	\includegraphics[width= 0.50\textwidth]{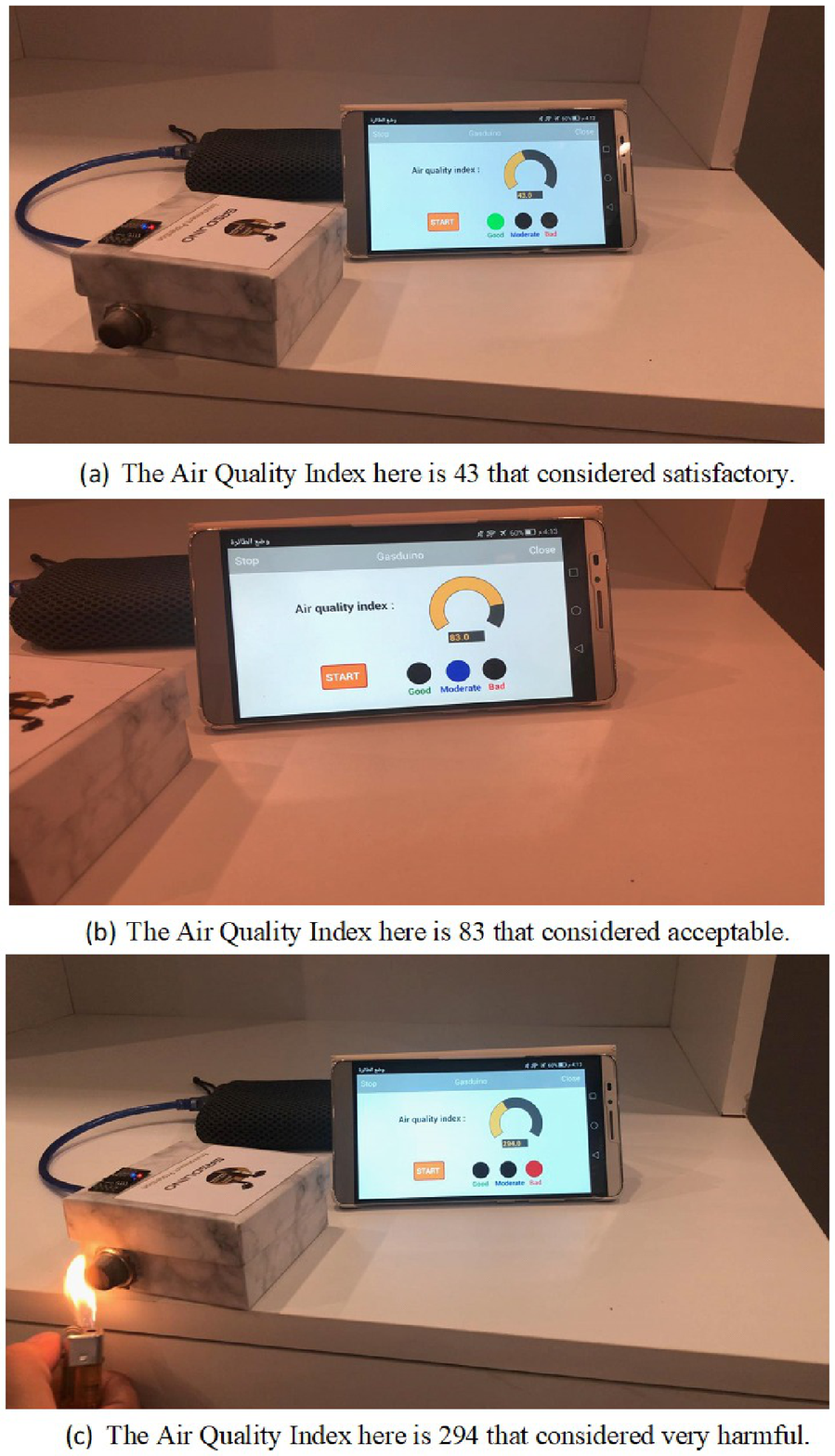}
	\caption{Practical tested cases of developed GASDUINO system.}
	\label{Fig.4}       
\end{figure*}

\section{Practical Results and Validation} \label{Sec2}
The operation of the developed GASDUINO system can be summarized as follows. First, the sensor (MQ-135) detects the gases in the air and analyzes them to estimate the AQI. Second, the AQI information is transmitted via the Wi-Fi model (ESP8266) to the IoT cloud, i.e. Remote XY clod. Finally, the developed Android UI shows the estimated value of AQI, and active the corresponding indicator, as shown in Figure \ref{Fig.2}.

The MQ-135 showed a fast response and accurate measurements of the air gases during this study. The developed GASDUINO system has been successfully tested to measure the air quality for two hours during the night and day, as depicted in Figure \ref{Fig.3}. The values of AQI are decreasing during the night from 70 to 40 PPM, while it begins to increase again from 30 to 44 PPM during the day. Moreover, the available three cases of the air quality monitoring and classification were successfully validated as shown Figure \ref{Fig.4}. Theses practical results demonstrated that our GASDUINO system can potentially be used a trusted air quality monitoring application, which is able to follow the standard categories of the air quality, as illustrated in Table \ref{Tab.1}.

\section{Conclusions and Future Work} \label{Sec3}
This study presented a new mobile application; namely GASDUINO to allow the users to measure and classify the level of the air pollution using the Internet of Things (IoT) successfully. The future work of this project will include the following prospects: 1) Create a network of wireless gas sensors to cover more than one place simultaneously for monitoring the AQI during 24 hours; 2) Improve the developed mobile application to display the percentage of each gas emission individually in the air with the overall PPM values; 3) Using artificial intelligence models like neural networks \cite{c18} and fuzzy systems \cite{c19} for predicting the air quality levels per time, e.g. weeks or months; and 4) Applying the developed GASDUINO system in the industrial areas of Shaqra city, Saudi Arabia.

\section*{Acknowledgment}
This project is financially supported by the College of Computing and Information Technology, Shaqra University.

\emergencystretch=\hsize

\begin{center}
\rule{6 cm}{0.02 cm}
\end{center}

\end{document}